\documentclass[a4paper,fleqn]{cas-dc}
\usepackage[numbers]{natbib}
\usepackage{enumitem}
\usepackage{float}
\usepackage{graphicx}
\graphicspath{ {./figure/} }
\usepackage[font=rm,justification=centering]{caption}
\def\tsc#1{\csdef{#1}{\textsc{\lowercase{#1}}\xspace}}
\tsc{WGM}
\tsc{QE}

\begin{document}
\let\WriteBookmarks\relax
\def\floatpagepagefraction{1}
\def\textpagefraction{.001}
\let\printorcid\relax 

\shorttitle{}    

\shortauthors{Y. Wang \emph{et al.}}

\title[mode = title]{Characterizing Social Media Behavior under Limited Observability: Retweet Link Prediction and Multimodal Intention Recognition}

\author{Yifan Wang$^1$, Chenwei Wang$^2$}


\address{$^1$South China University of Technology, Guangzhou 511400, China (email: 1441297934@qq.com)\\
$^2$Google LLC, Mountain View, California 94043, US (email: chen.516@gmail.com)}

\begin{abstract}
 Social-media behavior can be observed through both interaction structure and multimodal content, yet each view is often incomplete. We study two complementary tasks on separate datasets: predicting retweet links from an extremely sparse graph without node attributes, and generating user intentions from text-image posts with limited annotations. For link prediction, we use a leakage-controlled transductive protocol to compare GCN and GraphSAGE, classical topology heuristics, and three node-input designs. Across repeated runs, a GCN with compact degree and strength features achieves an average precision of 0.9243 and an MCC of 0.7051. Under joint edge-label corruption, topology-based GCNs retain strong ranking performance, whereas fixed-threshold MCC degrades substantially; adding fixed random node identifiers does not improve overall robustness. For intention recognition, we retain the MIKO framework \cite{lu2024miko} and evaluate LoRA adaptation of three language models together with distribution-aware sampling based on Pearson correlation, cosine similarity, and KL divergence. Fine-tuning consistently improves lexical-overlap metrics, while distribution-aware sampling produces more uniform performance across intention types. Together, the two studies characterize who interacts and why users post when social-media observations are sparse, noisy, or limited.
\end{abstract}


\maketitle

\section{Introduction}
Social media platforms record user behavior in two complementary forms. Retweets, replies, and other interactions form a graph that reveals who engages with whom, while the text and images within posts provide evidence about why users communicate. In practice, neither view is complete: interaction graphs are extremely sparse and may contain recording errors, whereas multimodal intention data are expensive to annotate and therefore limited in scale.

Link prediction and intention recognition correspond to these structural and semantic views. The former asks whether an interaction between two users is plausible from the observed network; the latter asks what motivation is expressed by a post. We do not assume that the two datasets describe the same users, and we do not optimize a joint model. Instead, we study the tasks in parallel to examine how established graph and language models behave when their available evidence is sparse, noisy, or imbalanced.

\subsection{The Problems}
In this paper, we investigate the following two highly relevant tasks in the social network highlighted above. 
\begin{enumerate}[leftmargin=*, label=(\arabic*)]
    \item \emph{\textbf{Retweet Network Link Prediction}}: 
    Accurately predicting retweet connections helps us better understand how information spreads and how users engage on social media platforms. We seek to model the social network, primarily by leveraging the topological features of the network to predict whether a retweet relationship exists between users, which is a binary classification task.
    
    Because the retweet graph contains no user attributes, we ask how much predictive information can be recovered from simple topology and whether adding fixed random node identifiers provides complementary information. We also evaluate robustness when edge labels and the message-passing graph are corrupted consistently, which better represents misrecorded or spurious interactions than changing supervision labels while leaving the graph intact.
    
    \item \emph{\textbf{Multimodal Intention Recognition on Social Media}}: 
    Users on the target social network mainly express and interact via multi-modal content, e.g., integrating text, image, which not only carries information but also reflects internal motivations. Identifying these intentions is critical for decoding user behavior: it captures the "why" behind actions beyond surface content, and without this clarity, understanding of interaction remains fragmented, limiting interpretation of communication patterns or behavior prediction.

    We will infer ten types of underlying intentions, as shown in Table \ref {tab:intentions}, behind users' multimodal social media content, where nine of them are derived from the ATOMIC knowledge graph, and the remaining one is an open domain intention. By gaining in-depth insight into user intentions, we aim to enhance our understanding of user behavior and communication patterns on social media platforms.
\end{enumerate}

\begin{table}[ht]
\centering
\small
\rmfamily
\renewcommand{\arraystretch}{1.3}
\caption{\rmfamily\centering Ten Predefined Intention Types }
\label{tab:intentions}
\begin{tabular}{@{}ll@{}}
\toprule
\textbf{Type} & \textbf{Description} \\
\midrule
xNeed   & User’s need \\
xIntent & User’s intent \\
xAttr   & User’s attribute \\
xEffect & Effect of user’s behavior \\
xReact  & User’s reaction \\
xWant   & User’s desire \\
oEffect & Effect on others \\
oReact  & Others’ reaction \\
oWant   & Others’ desire \\
Open    & Open-domain motivation or purpose \\
\bottomrule
\end{tabular}
\end{table}

\subsection{Datasets}

A competition held at \emph{2024 IEEE Intl. Conf. on Big Data} \cite{sn_competition_2024} aimed to advance rigorous research on the characterization of user behavior on social networks through three key tasks. Specifically, the datasets associated with Task 1 and Task 3 align well with the scope of our investigation, thus forming the empirical foundation of our work in this paper. For this paper to be self-contained, we in this section briefly describe the datasets originally offered by \cite{sn_competition_2024}.

\subsubsection{Dataset of Retweet Network Link Prediction}

This dataset consists of a large-scale retweet network involving 1,563,421 users and 1,113,811 observed retweet edges, where each node represents a user, each edge denotes a retweet relationship, and the edge weight reflects the number of retweets. No additional user attributes or content features are included in this dataset.

In this paper, an edge $(u, v)$ from the original network is called positive if there is at least one recorded retweet interaction between the user $u$ and the user $v$. These edges, forming the positive population, are labeled $1$ and retain their original edge weight, which is the total count of retweets between them. Similarly, an edge $(u, v)$ is called negative if there is no recorded retweet interaction between user $u$ and user $v$. These edges, forming the negative population, are labeled $0$ and assigned a weight of $0$.

Since the network is highly sparse, the number of unobserved node pairs is much larger than the number of recorded edges. We first split observed positive edges and then sample non-edges independently within the training, validation, and test partitions. Sampled non-edges are checked against all recorded positive edges. They remain operational negatives rather than verified non-interactions, because an unrecorded pair may still contain an unobserved interaction.

With this dataset, we design and train models to predict whether unknown retweet relationships exist between nodes (label 0 or 1), while testing the model's robustness by introducing noise into the training data.

\subsubsection{Dataset of Multimodal Intention Recognition}

The multimodal social intention dataset consists of 979 Twitter posts with a total of 7,868 annotated intentions.
We randomly split the dataset into 881 training posts and a held-out set of 98 test posts. The 100-instance fine-tuning subsets described below are selected only from the training partition; the test set is kept fixed for the reported comparisons.

Table \ref{tab:intentions} summarizes the ten predefined intention types. In the ATOMIC-derived labels, ``x'' refers to the post author and ``o'' refers to other people affected by or responding to the author's behavior. The additional ``Open'' category captures motivations or purposes that are not restricted to the nine predefined ATOMIC relations.

Five annotators assessed the candidate intentions generated in relation to the original text-image pairs. Each intention was annotated with binary integers, thus forming a 10-dimensional binary vector represented by "intention\_labels". For example, if the sixth element intention\_labels[6] \(=1\)" (index starts from 1), then Intention 6 (oEffect) is the true intention of that tweet. Its corresponding natural language description might be "hoping other users become aware of the emergency at the airport after reading the tweet".

The core task for this dataset is to generate descriptive texts for each of the ten intention types based on the user's multi-modal content (e.g., text and image). To evaluate the quality of these generated intentions, we employ multiple metrics, including BLEU (with BLEU-4 as the key indicator), ROUGE-L, and BERTScore, which collectively measure the semantic similarity and linguistic quality between the generated descriptions and the reference annotations.

\subsection{Our Contributions}
In this paper, we investigate the two aforementioned tasks and summarize our main contributions as follows.

\begin{enumerate}[leftmargin=*, label=(\arabic*)]
\item \emph{\textbf{Leakage-controlled Retweet Link Prediction}}:
We construct a transductive evaluation protocol that splits observed positive edges before negative sampling, keeps validation and test targets outside the message-passing graph, and repeats model comparisons over ten random seeds. We compare GCN \cite{kipf2016semi}, GraphSAGE \cite{hamilton2017inductive}, an attention-based GATv2 baseline \cite{brody2022attentive}, and classical topology heuristics.

\item \emph{\textbf{Topology and Corruption Analysis}}:
We compare repeated node-strength features, compact degree-strength features, and topology concatenated with fixed random node identifiers. A joint corruption protocol changes observed training labels, the message graph, and topology-derived features consistently. The results separate ranking robustness, measured by average precision, from fixed-threshold decision robustness, measured by MCC.

\item \emph{\textbf{Small-sample Multimodal Intention Generation}}:
Building on MIKO \cite{lu2024miko}, we evaluate LoRA fine-tuning of Llama-3, Qwen2.5, and InternLM2.5 and compare three distribution-aware sampling criteria. Fine-tuning improves the observed generation metrics, while Pearson-based sampling yields the most uniform BERTScore across the ten intention types in our experiments.
\end{enumerate}

\section{Preliminaries}
\subsection{Model Approaches for Link Prediction in Social Networks}

Link prediction in social networks aims to infer potential connections based on the observed network structure (nodes and edges), which is a core task for understanding user interaction patterns and information diffusion. Existing methods can be categorized into three groups: traditional topology-based methods, embedding-based representation learning, and deep models based on GNNs. In this section, we briefly introduce each group.

First, traditional topology-based methods extract local structural features of node pairs for prediction, such as the number of common neighbors, the Jaccard coefficient, and the Adamic-Adar index. They are computationally simple and interpretable, but struggle to capture global dependencies and cannot deal with complex social networks well. 

Second, embedding-based methods, such as DeepWalk \cite{perozzi2014deepwalk} and Node2Vec \cite{grover2016node2vec}, generate node sequences via random walks and learn node embeddings using Word2Vec-like models, converting link prediction into vector similarity computation. These methods can effectively encode structural roles, but suffer from a disconnection between embedding learning and downstream tasks, thus being less robust to sparse networks.

Finally, GNNs in recent years have become the main framework for link prediction due to their end-to-end learning capabilities. Among them, GCNs \cite{kipf2016semi} realize local message passing through node neighborhoods through spectral graph theory, integrating both node features and topology, and has improved performance of transductive tasks in dense networks. 
However, GCNs rely on the full graph structure, thus difficult to scale to large-scale dynamic social networks and having weak generalization to unseen nodes. To address this, GraphSAGE \cite{hamilton2017inductive} leverages an inductive learning framework that samples fixed-size neighborhoods and learns generic aggregation functions (e.g., mean, max, LSTM aggregation). This enables effective modeling of new nodes and dynamic networks, and better aligns with the evolving nature of real-world social networks.

Note that in the presence of noisy or incomplete graph data, prior work has also explored techniques such as edge dropout, feature perturbation, and robust loss functions to enhance the noise resilience and stability of GNNs. In this paper, with these insights, we provide a comparative study of GCN and GraphSAGE under controlled noise and focus on their robustness and adaptability in sparse social graphs.

\subsection{Multimodal Intention Recognition and Large Model Adaptation}
\label{sec:multimodal_background}

Multimodal understanding has advanced with vision-language pretraining (VLP), which enables joint modeling of visual and textual information. While earlier methods often relied on object detection before downstream processing, approaches such as ViLT directly encode visual tokens without explicit region proposals \cite{gonella2023visual}. Contrastive strategies such as CLIP further improve cross-modal alignment and support zero-shot transfer.

Efforts such as Dai \emph{et al.} in \cite{dai2022enabling} and Fang \emph{et al.} in \cite{fang2023knowledge} explored cross-modal knowledge distillation techniques, allowing models like CLIP to gain generative capabilities while preserving strong modality alignment. Recent surveys \cite{huang2024survey, Lymperaiou2024survey} have emphasized the role of structured knowledge in enhancing reasoning and interpretability in multimodal systems, revealing a growing trend toward building models capable of deeper understanding rather than simple pattern matching.

To further enhance the semantic grounding of multimodal systems, knowledge-enhanced learning has been proposed. For example, Lu \emph{et al.} introduced the MIKO framework in \cite{lu2024miko}, which distills common sense knowledge of social media from large language models to guide multimodal intention modeling. This approach bridges the gap between visual perception and high-level intent understanding by injecting curated external knowledge.

These developments underscore a shift toward more integrated, knowledge-aware, and interpretable multimodal systems, laying the foundation for robust user behavior modeling in social media environments.

\section{Methodology for Retweet Network Link Prediction}

\begin{figure*}[!t]
    \centering
    \includegraphics[width=6.8in]{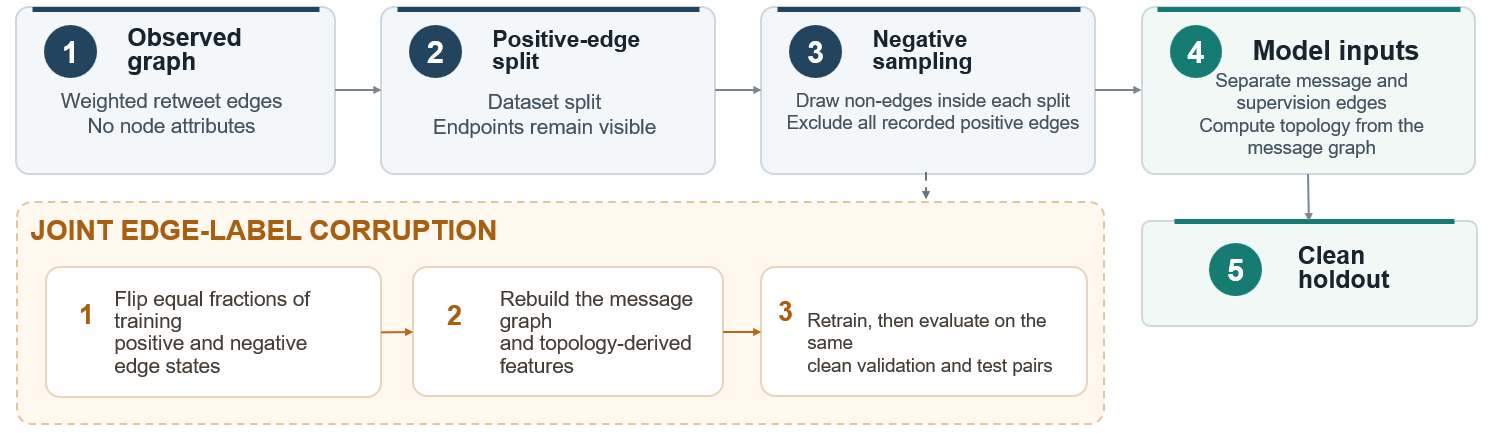}
    \caption{\rmfamily\centering Link-prediction protocol. Positive edges are split before split-specific negative sampling. Message-passing edges are disjoint from supervised and evaluation pairs. Joint corruption rebuilds both the graph and topology-derived node inputs before evaluation on a clean holdout set.}
    \label{scheme_link_prediction}
\end{figure*}

The first task predicts whether an unobserved retweet link is plausible between two users in a large-scale undirected graph. Nodes represent users, weighted edges represent recorded retweet relationships, and no user-level attributes or textual features are available. Figure \ref{scheme_link_prediction} summarizes the leakage-controlled split, feature construction, clean evaluation, and joint corruption protocol.

\subsection{Data Preprocessing}

The original dataset is represented as a highly sparse undirected network with $N=1,563,421$ nodes and $E=1,113,811$ recorded edges. Its density is $\rho = \frac{2E}{N(N-1)}\approx 9.11 \times 10^{-7}$. We canonicalize each undirected pair and merge duplicate observations while retaining the aggregate retweet weight.



We first assign recorded positive edges to training, validation, and test partitions. The transductive split preserves endpoint visibility: every endpoint of a validation or test edge also occurs in the training graph. Within the outer training partition, a second edge split creates a message-passing graph and a disjoint supervised set, preventing a target edge from being used simultaneously as a message edge. Only after these positive-edge splits do we independently sample node pairs that are absent from the complete recorded edge set. Positives and sampled negatives are balanced at 1:1 within each supervised or evaluation partition. This ratio defines a controlled classification benchmark and does not reproduce the prevalence of all possible non-edges in the original graph.

\begin{table}[ht]
\centering
\small
\rmfamily
\renewcommand{\arraystretch}{1.3}
\caption{\rmfamily\centering Link-prediction evaluation protocol}
\label{table ratio}
\begin{tabular}{@{}p{4.2cm}p{3.5cm}@{}}
\toprule
\textbf{Setting} & \textbf{Value} \\
\midrule
Outer positive-edge split & 98\% train, 1\% validation, 1\% test \\
Training supervision & 2\% of outer training positives \\
Positive-to-negative ratio & 1:1 in each partition \\
Clean repeated runs & 10 seeds (42--51) \\
Joint-corruption runs & 5 seeds (42--46) \\
\bottomrule
\end{tabular}
\end{table}

\subsection{Graph Models}
We compare two representative GNN architectures, GCN and GraphSAGE, using the same two-layer encoder and dot-product edge decoder. Hyperparameters are selected separately for each architecture using validation data rather than transferring the setting of one model to the other. We additionally report GATv2 as an attention-based clean-data baseline.

\begin{table}[h]
\small
\rmfamily
\centering
\caption{\rmfamily\bfseries\centering  GNN Model Architectures in Our Work}
\label{tab:model_compact}
\begin{tabular}{@{}p{1.5cm}>{\centering\arraybackslash}p{3.2cm}>{\centering\arraybackslash}p{2.8cm}@{}}
\toprule
\textbf{Part} & \textbf{GCN} & \textbf{GraphSAGE} \\
\midrule
Layer 1 & GCN($d\rightarrow64$) & SAGE($d\rightarrow64$) \\
Activation & ReLU & ReLU \\
Layer 2 & GCN(64→32) & SAGE(64→32) \\
Sample & -- & $[20,20]$ neighbors \\ 
Aggregation & Degree-normalized Sum & Mean  \\
Loss & BCEWithLogitsLoss & BCEWithLogitsLoss \\
\bottomrule
\end{tabular}
\end{table}

As summarized in Table \ref{tab:model_compact}, both encoders contain two message-passing layers. The first maps the input dimension $d$ to 64 hidden units, followed by ReLU and dropout, and the second produces a 32-dimensional node embedding. GCN aggregates the full message-passing neighborhood using degree-normalized summation. GraphSAGE instead applies mean aggregation to at most 20 sampled neighbors at each layer. For a candidate pair $(u,v)$, both architectures use the same parameter-free dot-product decoder,
\begin{equation}
z_{uv}=\mathbf z_u^{\top}\mathbf z_v,
\qquad
\widehat p_{uv}=\sigma(z_{uv}),
\end{equation}
where $\mathbf z_u$ and $\mathbf z_v$ are the final node embeddings. Model parameters are optimized with binary cross-entropy with logits on the supervised training pairs. Keeping the decoder, embedding size, loss, and evaluation protocol fixed isolates the effect of the two neighborhood-aggregation mechanisms.
\subsection{Feature Construction}
The retweet dataset provides edge weights but no user profiles, text embeddings, or other node attributes. A GNN nevertheless requires an input vector for every node. We therefore construct node inputs from two sources: activity statistics computed from the message-passing graph and fixed random vectors used as identifier-like controls. Importantly, validation and test target edges are excluded before these features are computed. Under joint corruption, the statistics are recomputed from the corrupted message graph rather than retained from the clean graph.

For a node $v$, let
\begin{equation}
d_v=|\mathcal{N}(v)|,
\qquad
s_v=\sum_{u\in\mathcal{N}(v)}w_{uv},
\end{equation}
where $d_v$ is the unweighted degree, $s_v$ is the weighted strength, and $w_{uv}$ is the recorded retweet count. Degree describes how many distinct users interact with $v$, whereas strength measures the total volume of those interactions. Because both quantities are highly skewed, we first apply a $\log(1+x)$ transformation and then scale each quantity by its maximum value on the message graph:
\begin{equation}
\widehat d_v=
\frac{\log(1+d_v)}{\max_j\log(1+d_j)},
\qquad
\widehat s_v=
\frac{\log(1+s_v)}{\max_j\log(1+s_j)}.
\end{equation}
An isolated node consequently receives $\widehat d_v=\widehat s_v=0$. We define the compact structural vector as $\mathbf c_v=[\widehat d_v,\widehat s_v]\in\mathbb{R}^{2}$ and the repeated-strength vector as
\begin{equation}
\mathbf t_v=\widehat s_v\mathbf 1_{128}\in\mathbb{R}^{128}.
\end{equation}
Repeating the same scalar does not introduce additional information; it preserves the 128-dimensional input convention of the original experiments and allows the effect of input dimension to be separated from the structural statistic itself.

For the random component, each node receives
\begin{equation}
\mathbf r_v\sim\mathcal{N}(\mathbf 0,\mathbf I_{128}).
\end{equation}
The vector is generated once from the run seed and then held fixed during optimization. It is therefore neither a trainable embedding nor fresh noise injected at every epoch. In a transductive split it can act as an arbitrary node identifier, but it carries no intrinsic social or structural meaning. This distinction is important: the experiments test whether fixed identifiers complement topology, not whether stochastic feature perturbation regularizes the model.

Using $\mathbf c_v$, $\mathbf t_v$, and $\mathbf r_v$, we evaluate the following six node-input configurations:
\begin{enumerate}[leftmargin=*, label=(\arabic*)]
    \item \textit{Repeated topology}: $\mathbf h_v=\mathbf t_v\in\mathbb{R}^{128}$. This is the topology-only construction retained from the original paper.

    \item \textit{Topology + fixed random IDs}: concatenate 64 repeated-strength entries with 64 random entries,
    \begin{equation*}
    \mathbf h_v=[\mathbf t_v[1{:}64];\mathbf r_v[1{:}64]]\in\mathbb{R}^{128}.
    \end{equation*}
    This corresponds to the original concatenation strategy.

    \item \textit{Weighted topology/random}: combine the two 128-dimensional vectors elementwise with equal weights,
    \begin{equation*}
    \mathbf h_v=0.5\mathbf t_v+0.5\mathbf r_v\in\mathbb{R}^{128}.
    \end{equation*}

    \item \textit{Interleaved topology/random}: alternate structural and random entries,
    \begin{equation*}
    h_v[i]=
    \begin{cases}
    t_v[i], & i \text{ is odd},\\
    r_v[i], & i \text{ is even}.
    \end{cases}
    \end{equation*}
    This is the corrected implementation of the original dimension-replacement strategy.

    \item \textit{Compact topology}: $\mathbf h_v=\mathbf c_v\in\mathbb{R}^{2}$. Unlike repeated topology, this representation retains both the number of distinct neighbors and total retweet volume without duplicating dimensions.

    \item \textit{Fixed random IDs}: $\mathbf h_v=\mathbf r_v\in\mathbb{R}^{128}$. This control measures how much the model can obtain from graph propagation plus arbitrary node identifiers when explicit activity features are removed.
\end{enumerate}

All six configurations are included in the clean GCN feature ablation in Section \ref{sec:feature_ablation}. The robustness experiment concentrates on repeated topology, compact topology, and topology with fixed random IDs: together, these compare the original structural representation, the most economical structural representation, and the principal topology-random fusion strategy.

\subsection{Joint Edge-Label Corruption}
Changing only supervised labels while leaving message passing and topology-derived features clean creates an inconsistent training graph. We therefore corrupt the observed training data before constructing the message graph. At corruption ratio $r$, we select $r$ of the original training positives and relabel them as absent, and select the same proportion of sampled training negatives and relabel them as observed. The resulting observed positives contain both retained true edges and injected false edges. We then create disjoint message and supervision subsets from this corrupted pool and recompute degree and strength from the corrupted message graph.

We evaluate $r\in\{0,5,10,\ldots,50\}\%$ for GCN and GraphSAGE, the three principal node-input designs, and five seeds. Validation and test pairs remain clean so that degradation is measured against unchanged ground truth. Early stopping uses validation average precision. For deployment-oriented comparison, the MCC decision threshold at each noisy condition is fixed to the threshold selected on the corresponding clean validation run. We report average precision (AP) for ranking quality and MCC for fixed-threshold classification, together with mean, standard deviation, and area under the robustness curve (AURC).

\section{Methodology for Multimodal Social Media Content Intention Recognition}

\subsection{The Proposed Solution}
We retain the core MIKO pipeline \cite{lu2024miko}: visual evidence is first converted into an image description, textual and visual cues are condensed into keywords, and an LLM generates natural-language descriptions for the applicable intention types. Our implementation keeps LLaVA-v1.5-7B for image description, replaces ChatGPT-based keyword extraction with DeepSeek-V3 while retaining the original prompts, and replaces the original Llama2-7B generator with a controlled comparison of LoRA-adapted Llama-3-8B, Qwen2.5-7B, and InternLM2.5-7B. We further compare random and distribution-aware selection of the 100-instance fine-tuning subset. Training settings are reported in Table \ref{table4}, and generation is evaluated using BLEU-4, ROUGE-L, and BERTScore F1.

\subsection{Data Processing}
\label{sec:intention_sampling}
The multimodal intention recognition dataset consists of 979 Twitter posts, where each post is structured as a quadruple: text content, accompanying image, 10-dimensional intention labels, and natural language descriptions corresponding to each intention type.

To implement the MIKO framework and our fine-tuning pipeline, we perform the following data processing steps:

\begin{enumerate}[leftmargin=*, label=(\arabic*)]
    \item \textbf{Image Processing}: First, we use the multimodal large model (LLaVA-v1.5-7B) to generate consistent and detailed descriptions for each post's image in the original dataset, covering core visual elements and potential social context clues to achieve cross-modal alignment.

    \item \textbf{Text Preprocessing}: We employ DeepSeek-V3 to extract keywords (including entities, action words, and emotional tendencies) from the original post-text and image descriptions, filtering redundant information to enhance the relevance between text and intention labels, serving the purposes of data cleaning and semantic focus \cite{lu2024miko}.

    \item \textbf{Distribution-aware Sampling for Training Data}: Because intention types are not equally frequent, we compare random selection with three strategies for selecting 100 training instances: maximizing Pearson correlation with the full training distribution, maximizing cosine similarity, and minimizing KL divergence. All candidate subsets are drawn only from the training partition; the held-out test partition is never used for subset selection. These criteria seek representative rather than artificially uniform label proportions.
\end{enumerate}

The above image processing, text preprocessing, and final intention generation steps all follow the core workflow of the MIKO framework. The processed data, structured as "original post text + image description + keywords + intention labels + reference descriptions", serves as input specifically for fine-tuning the LLM responsible for the final intention description generation component within the framework.

\section{Experiments for Retweet Network Link Prediction}

\begin{table}[ht]
\centering
\small
\rmfamily
\renewcommand{\arraystretch}{1.3}
\caption{\rmfamily\centering Training Hyperparameters for Link Prediction Models}
\label{table strategy}
\begin{tabular}{@{}p{3.3cm} >{\centering\arraybackslash}p{2.2cm} >{\centering\arraybackslash}p{2.2cm}@{}}
\toprule
\textbf{Hyperparameter} & \textbf{GCN} & \textbf{GraphSAGE} \\
\midrule
Optimizer & Adam & Adam \\
Learning rate & $5\times10^{-3}$ & $5\times10^{-4}$ \\
Weight decay & $10^{-4}$ & $10^{-4}$ \\
Dropout & 0.1 & 0.2 \\
Mini-batch size & 8192 & 8192 \\
Maximum epochs & 2000 & 2000 \\
Minimum epochs / patience & 30 / 20 & 30 / 20 \\
Early-stopping metric & Validation AP & Validation AP \\
Prediction threshold & \multicolumn{2}{c}{Selected on validation MCC} \\
\bottomrule
\end{tabular}
\end{table}

\begin{table}[ht]
\centering
\small
\rmfamily
\renewcommand{\arraystretch}{1.3}
\caption{\rmfamily\centering Fine-tuning Hyperparameters for Intention Generation}
\label{table4}
\begin{tabular}{@{}p{3.5cm} >{\centering\arraybackslash}p{3.5cm}@{}}
\toprule
\textbf{Hyperparameter} & \textbf{Value} \\
\midrule
Trainable Layers & Last 2 layers \\
Epochs & 10 \\
Initial Learning Rate & 5e-5 \\
Optimizer & AdamW \\
Batch Size & 2 \\
LoRA Rank & 8 \\
LoRA Scaling Factor & 32 \\
\bottomrule
\end{tabular}
\end{table}


\begin{table*}[!t]
\small
\rmfamily
\centering
\caption{\rmfamily\centering Transductive link-prediction performance over 10 seeds (mean $\pm$ standard deviation).}
\label{tab:gnn_comparison}
\begin{tabular}{@{}llccc@{}}
\toprule
\textbf{Method} & \textbf{Node input} & \textbf{ROC-AUC} & \textbf{AP} & \textbf{MCC} \\
\midrule
Preferential Attachment & -- & $0.8733\pm0.0016$ & $0.8708\pm0.0015$ & $0.6838\pm0.0049$ \\
GraphSAGE & Repeated topology & $0.9157\pm0.0019$ & $0.9120\pm0.0017$ & $0.6795\pm0.0054$ \\
GraphSAGE & Compact topology & $0.9122\pm0.0020$ & $0.9117\pm0.0018$ & $0.6796\pm0.0041$ \\
GATv2 & Repeated topology & $0.8778\pm0.0141$ & $0.8999\pm0.0049$ & $0.6755\pm0.0098$ \\
GCN & Repeated topology & $0.9260\pm0.0018$ & $0.9212\pm0.0011$ & $0.7000\pm0.0033$ \\
GCN & Compact topology & $\mathbf{0.9295\pm0.0016}$ & $\mathbf{0.9243\pm0.0016}$ & $\mathbf{0.7051\pm0.0038}$ \\
\bottomrule
\end{tabular}
\end{table*}

\subsection{Link Prediction Performance}

Table \ref{tab:gnn_comparison} reports the clean transductive results. Unlike our initial single-configuration comparison, each GNN is tuned separately using validation data and evaluated over ten seeds. GCN is the strongest architecture in this setting: with compact topology it reaches AP $0.9243\pm0.0016$ and MCC $0.7051\pm0.0038$. Repeated topology gives a closely comparable result. GraphSAGE and GATv2 remain competitive, but neither improves on the tuned GCN. Preferential Attachment is a strong non-neural baseline, confirming that node activity is highly informative in this retweet graph. We therefore use GCN as the principal model in the robustness analysis and retain GraphSAGE as an architecture-sensitivity control.

\begin{figure*}[!t]
\centering
\includegraphics[width=6.8in]{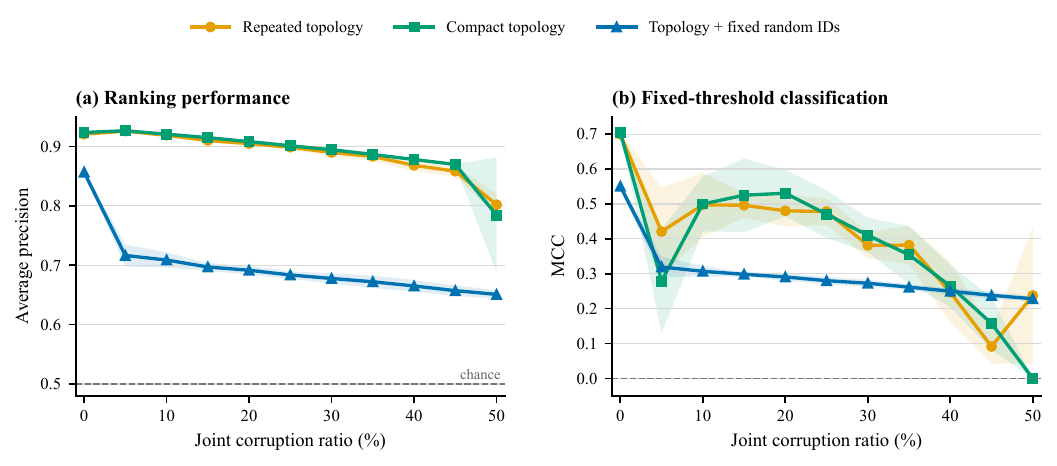}
\caption{\rmfamily\centering GCN performance under joint edge-label corruption. Curves show the mean over five seeds and shaded regions show one standard deviation. AP measures ranking performance; MCC uses the threshold selected on the corresponding clean validation run.}
\label{fig:robust}
\end{figure*}

\begin{table*}[!t]
\small
\rmfamily
\centering
\caption{\rmfamily\centering GCN feature ablation over 10 transductive seeds (mean $\pm$ standard deviation).}
\label{tab:strategy_comparison}
\begin{tabular}{@{}lccc@{}}
\toprule
\textbf{Node input} & \textbf{Dimension} & \textbf{AP} & \textbf{MCC} \\
\midrule
Repeated topology & 128 & $0.9212\pm0.0011$ & $0.7000\pm0.0033$ \\
Topology + fixed random IDs & 128 & $0.8567\pm0.0024$ & $0.5509\pm0.0075$ \\
Weighted topology/random & 128 & $0.8412\pm0.0019$ & $0.5190\pm0.0057$ \\
Interleaved topology/random & 128 & $0.8566\pm0.0038$ & $0.5497\pm0.0058$ \\
Compact topology & 2 & $\mathbf{0.9243\pm0.0016}$ & $\mathbf{0.7051\pm0.0038}$ \\
Fixed random IDs & 128 & $0.7838\pm0.0053$ & $0.4384\pm0.0071$ \\
\bottomrule
\end{tabular}
\end{table*}
\subsection{Comparison of Feature Processing Strategies}
\label{sec:feature_ablation}

Table \ref{tab:strategy_comparison} shows that the graph's simple activity statistics contain most of the usable signal. Compact topology uses only degree and weighted strength yet slightly exceeds the 128-dimensional repeated-strength input. Because the difference is small, we interpret compact topology as a more economical and interpretable representation rather than as a statistically established accuracy improvement.

Adding fixed random node identifiers consistently reduces both AP and MCC. Weighted and interleaved combinations show the same pattern, and fixed random identifiers alone perform worst. These results do not support the hypothesis that static random identifiers provide useful complementary information. They also should not be interpreted as a test of stochastic feature-noise regularization: the random identifiers are sampled once and remain fixed.

\subsection{Performance Under Noisy Conditions}

Figure \ref{fig:robust} reveals different behavior for ranking and fixed-threshold classification. The AP of the two topology-based GCNs decreases gradually: repeated topology remains at 0.8682 under 40\% corruption and 0.8019 under 50\%, while compact topology reaches 0.8780 and 0.7842. In contrast, topology with fixed random identifiers drops from a clean AP of 0.8569 to 0.7164 after only 5\% corruption and reaches 0.6509 at 50\%. Its AP-AURC is 0.6923, compared with 0.8919 for repeated topology and 0.8954 for compact topology.

MCC declines much faster and is non-monotonic for the topology-based inputs. This is primarily a threshold-calibration effect rather than an immediate loss of ranking information. For example, repeated topology at 5\% corruption has AP 0.9258 but MCC 0.4208: mean precision is 0.9808 while recall falls to 0.3249 under the clean validation threshold. At 45\% corruption, AP remains 0.8584, but recall is only 0.0205 and MCC is 0.0915. Thus, joint corruption shifts the score distribution enough to invalidate a fixed clean threshold even while many positive links remain correctly ranked.

Across the five paired seeds, fixed random identifiers have lower GCN AP-AURC and MCC-AURC than repeated topology for every seed. The corresponding two-sided exact Wilcoxon tests give $p=0.0625$, the smallest attainable nonzero two-sided value with five nonzero pairs. We therefore report a consistent direction and effect size, but do not claim significance at the 0.05 level. GraphSAGE shows substantially larger degradation and is included as an architecture-sensitivity result rather than evidence of a universal GNN response.

\renewcommand{\baselinestretch}{1.2}
\begin{table*}[ht]
    \small
    \rmfamily
    \centering
    \caption{\rmfamily\centering Intention-generation results for base and LoRA-adapted models and for Llama-3 trained with different subset-selection criteria. ``FT'' denotes LoRA fine-tuning; models in the upper block use the same randomly sampled training subset.}
    \label{tab:intention_generation}
    \begin{tabular}{@{}p{4.2cm}cccccccc@{}}
        \toprule
        \textbf{Model} & \textbf{BLEU-4} & \textbf{ROUGE-1} & \textbf{ROUGE-2} & \textbf{ROUGE-L} & \textbf{BERTScore-P} & \textbf{BERTScore-R} & \textbf{BERTScore-F1} \\ \midrule
        Llama-3-8B-Instruct & 39.9742 & 49.5160 & 29.8204 & 38.7724 & 0.900 & 0.894 & 0.897 \\
        FT\_Llama-3-8B-Instruct & 62.8274 & \underline{62.6331} & \underline{48.0902} & \textbf{56.5071} & \textbf{0.903} & \textbf{0.905} & \textbf{0.904} \\
        Qwen2.5-7B-Instruct & 58.9060 & 57.6028 & 36.1127 & 50.3438 & 0.901 & 0.904 & 0.903 \\
        FT\_Qwen2.5-7B-Instruct & \underline{63.3092} & 61.8879 & 47.8305 & \underline{56.4084} & 0.901 & 0.904 & 0.903 \\
        InternLM2.5-7B-Chat & 52.6799 & 55.0325 & 34.1849 & 47.8048 & 0.901 & 0.896 & 0.898 \\
        FT\_InternLM2.5-7B-Chat & \textbf{63.3976} & \textbf{62.9002} & \textbf{48.7041} & 56.1197 & 0.899 & 0.901 & 0.900 \\ \midrule
        Pearson Sampling(FT\_Llama) & 64.8939 & 65.3064 & 52.3193 & 60.126 & 0.893 & 0.91 & 0.902 \\
        CS Sampling(FT\_Llama) & 63.4322 & 62.9889 & 48.641 & 57.3925 & 0.906 & 0.889 & 0.897 \\
        KL Sampling(FT\_Llama) & 63.2451 & 62.6114 & 48.7676 & 57.418 & 0.907 & 0.89 & 0.898 \\ \bottomrule
    \end{tabular}
\end{table*}

\begin{table*}[ht]
    \small
    \rmfamily
    \centering
    \caption{\rmfamily\centering Mean BERTScore by intention type for Llama-3 fine-tuned with different subset-selection criteria.}
    \label{tab:intention_by_type}
    \begin{tabular}{@{}p{1.1cm}cccccccccccc@{}}
        \toprule
        \textbf{Standard} & \textbf{xWant} & \textbf{oEffect} & \textbf{xAttr} & \textbf{xIntent} & \textbf{xReact} & \textbf{oReact} & \textbf{oWant} & \textbf{xEffect} & \textbf{xNeed} & \textbf{Open} & \textbf{Average} & \textbf{Std. Dev.} \\ \midrule
        Random & 0.903 & \textbf{0.916} & 0.903 & 0.89 & 0.897 & \textbf{0.907} & \textbf{0.914} & \textbf{0.903} & 0.885 & 0.893 & 0.9011 & 0.0095 \\
        Pearson & \textbf{0.918} & \underline{0.914} & \textbf{0.913} & \textbf{0.912} & \textbf{0.906} & \underline{0.905} & \underline{0.905} & \textbf{0.903} & \textbf{0.902} & \textbf{0.902} & \textbf{0.9080} & \textbf{0.0054} \\
        CS & \underline{0.917} & 0.911 & 0.91 & 0.908 & 0.903 & 0.902 & 0.902 & \underline{0.9} & 0.8988 & 0.897 & 0.9048 & 0.0061 \\
        KL & 0.916 & 0.913 & \underline{0.911} & \underline{0.91} & \underline{0.905} & \underline{0.905} & 0.904 & \underline{0.9} & \underline{0.9} & \underline{0.899} & \underline{0.9064} & \underline{0.0056} \\ \bottomrule
    \end{tabular}
\end{table*}

\begin{table*}[ht]
\centering
\small
\rmfamily
\caption{\rmfamily\centering Results of Sampled Intention Distributions Under Different Metrics}
\label{table5}
\begin{tabular}{@{}lcccccccccc@{}}
\toprule
\textbf{Metric} & \textbf{xNeed} & \textbf{xIntent} & \textbf{xAttr} & \textbf{xEffect} & \textbf{xReact} & \textbf{xWant} & \textbf{oEffect} & \textbf{oReact} & \textbf{oWant} & \textbf{Open} \\
\midrule
Expected value & 82 & 81 & 78 & 78 & 63 & 74 & 79 & 74 & 69 & 80 \\
Best Pearson & 85 & 83 & 79 & 80 & 58 & 73 & 80 & 73 & 64 & 83 \\
Best Cosine & 84 & 80 & 77 & 78 & 63 & 73 & 81 & 74 & 69 & 80 \\
Min KL Divergence & 82 & 82 & 79 & 80 & 65 & 76 & 81 & 74 & 69 & 82 \\
\bottomrule
\end{tabular}

\vspace{0.5em}
\rmfamily\small
Note: Best Pearson correlation = 0.995, Best Cosine similarity = 0.9983, Minimum KL divergence = 0.000067.
\end{table*}

\section{Experiments for Multimodal Social Media Content Intention Recognition}

We adopt the MIKO framework \cite{lu2024miko} for data processing, replacing ChatGPT with DeepSeek-V3 \cite{liu2024deepseek} for keyword extraction while retaining the original prompts. We fine-tune Llama-3-8B-Instruct \cite{touvron2024llama3}, Qwen2.5-7B-Instruct \cite{team2024qwen2}, and InternLM2.5-7B-Chat \cite{cai2024internlm2} for intention generation using LoRA. Hyperparameters are summarized in Table \ref{table4}, and evaluation uses BLEU-4, ROUGE, and BERTScore.

From the 881-instance training partition, we draw 10,000 candidate subsets of 100 instances and select one subset under each distribution criterion. The 98-instance test partition remains fixed across all comparisons. Table \ref{table5} reports the selected label counts, and Table \ref{tab:intention_by_type} reports BERTScore by intention type.

\subsection{Model Fine-tuning Effects and Performance Comparison}
The upper part of Table \ref{tab:intention_generation} compares each base model with its LoRA-adapted counterpart on the same held-out set. All three adapted models improve substantially on BLEU and ROUGE. BERTScore changes are smaller: Llama-3 increases from 0.897 to 0.904, Qwen2.5 remains at 0.903, and InternLM changes from 0.898 to 0.900. Because these values come from one data split without repeated-run uncertainty, we describe the observed differences without a statistical-significance claim.

Differences among the adapted models are modest and metric dependent. Llama-3 has the highest BERTScore F1 and nearly the highest overlap scores, so we use it as the common backbone for the sampling comparison.

\subsection{Impact of Sampling Strategies on Intention Generation}
Random sampling can yield an unrepresentative mixture of intention types. We therefore compare the Pearson, cosine-similarity, and KL-divergence criteria described in Section \ref{sec:intention_sampling}. The results are shown in the lower part of Table \ref{tab:intention_generation} and in Table \ref{tab:intention_by_type}.

The aggregate scores differ only modestly across selection criteria. Pearson selection gives the highest reported overlap scores and raises the mean per-intention BERTScore from 0.9011 to 0.9080 while reducing its across-intention standard deviation from 0.0095 to 0.0054. Cosine and KL selection show the same lower-variance direction. These descriptive results suggest that matching the training-label distribution can improve uniformity across intention types, but repeated data splits are needed before claiming general superiority.

\section{Conclusion}

This paper studies two complementary views of social-media behavior under limited observability: interaction structure for retweet link prediction and multimodal content for intention generation. Under a leakage-controlled transductive protocol, a tuned GCN with compact degree and strength features is the strongest evaluated link predictor, while a competitive preferential-attachment baseline shows how much signal is already present in activity patterns. Fixed random node identifiers do not improve accuracy or robustness. When corrupted edges alter both supervision and message passing, topology-based GCNs preserve useful ranking quality longer than fixed-threshold decisions, exposing calibration as a separate failure mode. In the content study, LoRA adaptation improves lexical-overlap metrics, and distribution-aware subset selection yields more even BERTScore across intention types on the retained split. These two tasks are evaluated separately rather than as a joint model. Future work should test inductive node splits, temporally grounded negatives, repeated multimodal splits, and genuinely coupled structural-content models before broader deployment claims are made.

\bibliographystyle{cas-model2-names}
\bibliography{cas-refs-arxiv-v1}
\end{document}